# Non-Contact and Non-Destructive Detection of Structural Defects in Bioprinted Constructs Using Video-Based Vibration Analysis

Md Anisur Rahman[1], Md Asif Hasan Khan[1], Tuan Mai[1], and Jinki Kim[1]*

[1] Department of Mechanical Engineering, Georgia Southern University, Statesboro, GA, USA 30458

* jinkikim@georgiasouthern.edu

**Abstract**

Bioprinting technology has advanced significantly in the fabrication of tissue-like constructs with complex geometries for regenerative medicine. However, maintaining the structural integrity of bioprinted materials remains a major challenge, primarily due to the frequent formation of hidden defects, such as pressure-induced inconsistencies, geometric voids, and interlayer separations. Traditional defect detection approaches often require physical contact that may not be suitable for hydrogel-based biomaterials due to the inherently soft nature of bioinks, making non-invasive and straightforward structural evaluation necessary in this field. To advance the state of the art, this study presents a novel non-contact method for non-destructively detecting structural defects in bioprinted constructs using video-based vibration analysis. Ear-shaped constructs were fabricated using a bioink composed of sodium alginate and κ-carrageenan using extrusion-based bioprinting. To simulate printing defects, controlled geometric, interlayer, and pressure-induced defects were systematically introduced into the samples. The dynamic response of each structure was recorded under excitation using a high-speed camera and analyzed using phase-based motion estimation techniques. Experimental results demonstrate that all defective samples exhibit consistent changes in the dynamic characteristics compared to baseline samples, with increasingly pronounced deviation observed as defect severity increases. These shifts reflect changes in effective stiffness and mass distribution induced by internal anomalies, even when such defects are not detectable through surface inspection. The experimental trends were also validated through finite element simulations. Overall, this work demonstrates that video-based vibrometry is a powerful approach for assessing the quality of bioprinted constructs, offering a practical pathway toward robust structural health monitoring in next-generation bio-additive manufacturing workflows.

1. **Introduction**

**1.1 Background**

3D bioprinting has been widely applied in the biomedical sector, enabling the fabrication of biomimetic and functional tissue constructs and organs [1], [2]. Its applications include the development of patient-specific implants [3], [4], microvasculature fabrication [5], disease modeling [6], scaffolds for tissue regeneration [7], [8], and fabricating organoids, vascularized constructs, and functional tissues [9]. Among various bioprinting techniques, extrusion-based bioprinting has been one of the most widely adopted approaches due to its adaptability, cost efficiency, and compatibility with a wide range of hydrogel-based bioinks [10], [11]. Despite the advances in material formulations and printer control, achieving high structural fidelity remains a fundamental challenge in extrusion-based bioprinting. The deposition process is sensitive to numerous factors, including the rheological properties of the bioink, suboptimal printing conditions [12], and unexpected events (such as printer nozzle clogging or pressure irregularities). These imperfections may appear as improper line width of the print, weak interlayer bonding, and discontinuities, which could severely compromise the functional and mechanical integrity of the printed construct and lead to catastrophic failures of bioprinted parts, including layer delamination, internal void formation, and filament collapse [13], [14], [15] [16], [17], [18], [19]. Monitoring and detection of these defects remains as one of the significant challenges in the widespread adoption of bioprinting. While several quality control strategies have been proposed, many rely on contact-based ultrasound sensors that may be incompatible with soft biomaterial [20], [21]. While it offers the advantage of internal imaging, its application in 3D bioprinting may be limited by difficulty in capturing fine structural discontinuities in hydrogel. Impedance-based monitoring is also adopted from civil and composite structures, which relies on analyzing changes in electrical properties due to mechanical anomalies [22], [23], [24], [25]. Its application in bioconstructs may be challenging by the requirement of embedded electrodes, which may not be ideal for soft and biocompatible systems. Optical coherence tomography-based monitoring approaches offer non-contact monitoring with micrometer-level resolution, making them available to assess internal structure in soft bioconstructs [26], [27], [28]. On the other hand, the effectiveness of these approaches can be limited in opaque or highly scattering bioinks where light penetration decreases. In addition, the cost and complexity of equipment and post processing may restrict accessibility [29], [30], [23], [24], [25]. To address these concerns and advance the state of the art, the new research presented in this article develops a novel method that enables a reliable, non-contact, and non-destructive method for detecting defects in bioprinted constructs. This study explores a new methodology that can evaluate the structural integrity of soft hydrogel-based bioprinted structures by measuring their dynamic response using video-based vibration analysis.

In the following sections, the bioink formulation and bioprinting setup employed to fabricate bioconstructs in this study are first introduced. Then a new approach for defect identification using video-based vibrometry is presented. Numerical and experimental investigations are conducted for various types of defects often observed in bioprinting to verify the effectiveness of the proposed method. Following these investigations, concluding remarks are provided to summarize the key findings and reflect upon the potential of the new approach.

## 2. Materials and Methods

### 2.1 Bioink Preparation

A bio-compatible double network hydrogel was formulated to serve as the structural material for fabricating bioconstructs used in this study. The hydrogel was composed of two naturally derived anionic polysaccharides, sodium alginate [31] and κ-carrageenan [32], each selected for its favorable gelation characteristics and suitability for extrusion-based bioprinting. The hydrogel was prepared by dissolving 3.7% (w/v) sodium alginate (Sigma-Aldrich) and 2% (w/v) κ-carrageenan (Kitchen Alchemy) in distilled water under continuous stirring at room temperature. Then 0.1 M homogeneous calcium chloride solution (Carolina) was gradually added (1.109% w/v) to initiate ionic crosslinking. This sequential mixing ensured uniform gelation and network integration. The final hydrogel mixture was placed in a vacuum chamber to remove air bubbles and ensure consistency in extrusion during bioprinting.

## 2.2 Bioprinting Setup

The fabrication of test samples in this study was performed using an extrusion-based bioprinter (INKREDIBLE, CELLINK). A tapered plastic nozzle with an inner diameter of 0.41 mm was employed to achieve fine deposition resolution. The extrusion pressure was maintained at 22 kPa using an integrated pneumatic regulator, ensuring consistent flow throughout the printing process. All samples were printed at room temperature onto glass Petri dishes. The printing was performed layer by layer following G-Code instructions generated via Cellink Heartware 2.4.1 software using a rectilinear infill pattern (50% infill density).

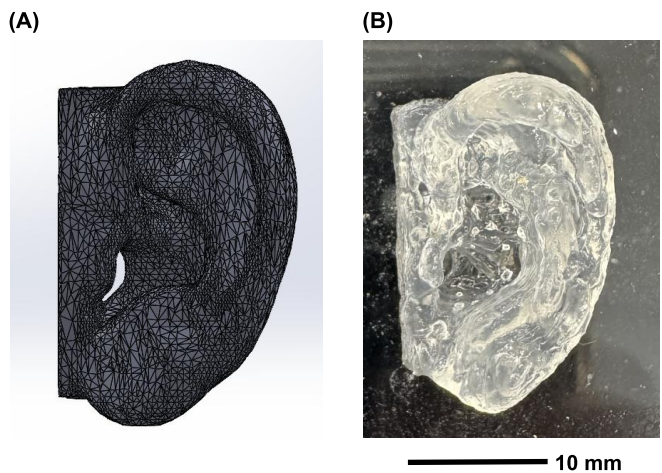

**Figure 1: Baseline ear structure: (A) CAD model, (B) Printed experimental model.**

Congenital ear conditions, such as microtia and trauma-induced ear deformities, are among the most reconstructed anatomical features using tissue engineering scaffolds. Inspired by the recent first substantial clinical success of printing and cultivating a patient's own cells for transplantation into a human organ, which was an ear [33], and considering its structural complexity and clinical relevance, a three-dimensional human ear model was employed as the target structural specimen for this study. The external ear presents a challenging geometry with varying curvature and wall thickness, making it ideal for evaluating the structural integrity in bioprinted constructs. Figure 1(A) shows the CAD model of ear structure used for baseline fabrication, while Figure 1(B) presents the corresponding 3D printed construct using the hydrogel formulation described in Section 2.1.

## 2.3 Video-based Vibration Measurement

The dynamic characteristics of the bioconstruct were investigated by recording its free vibration response using a high-speed camera (Chronos 1.4, Kron Technologies). Figure 2 represents the experimental setup for this method. Free vibration of the bioconstruct was induced by lifting one side of the Petri dish to a height of 2 mm and releasing it to apply a weak impulsive base excitation. Each recording was taken at a resolution of 1280×1024 pixels and a frame rate of 500 frames per second to capture subtle vibrational responses while maintaining high spatial fidelity. Uniform illumination was provided using external light sources without flickering to minimize shadowing and reflection from the translucent hydrogel surface.

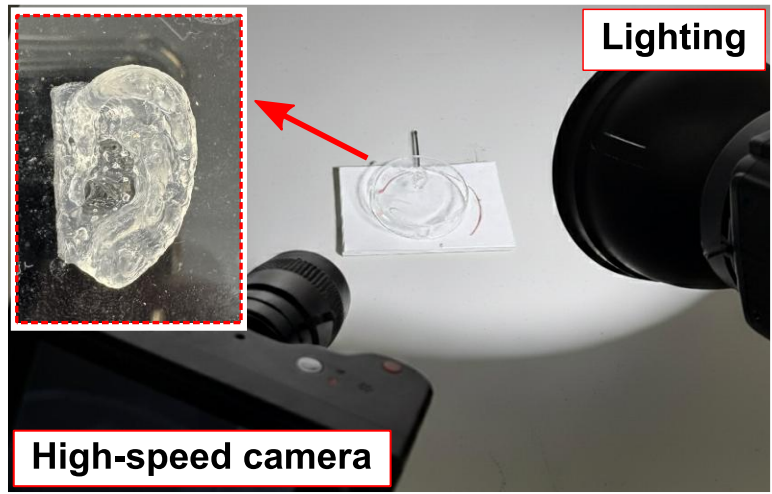

**Figure 2. Experimental Setup with high-speed camera, and external lighting.**

## 2.4 Video-Based Vibration Analysis

To assess the dynamic behavior of the bioprinted constructs, a video-based vibration analysis was employed. This approach relies on the phase-based motion estimation technique, which is capable of detecting minute vibration through subtle variations in pixel intensity over time [34]. When captured under uniform lighting conditions, temporal changes in pixel brightness are intrinsically linked to the structural motion of the object, allowing the extraction of vibrational characteristics. To overcome the limitation of conventional Fourier analysis for localized movement in soft bio constructs, two-dimensional Gabor wavelet filters (shown in Figure 3) have been employed. These filters enhance the sensitivity of the motion estimation to spatially localized displacement. Moreover, they decompose the video frames along horizontal and vertical orientations. Each pixel intensity signal, $I(x, y, t)$, in the video sequence is convolved with a complex Gabor kernel composed of real and imaginary components. This operation yields localized amplitude $A_\theta(x, y, t)$ and phase $\phi_\theta(x, y, t)$ terms as:

$$\left(G_2^\theta + iH_2^\theta\right) \otimes I(x, y, t) = A_\theta(x, y, t) e^{i\phi_\theta(x,y,t)} \tag{1}$$

where $G_2^\theta$ and $H_2^\theta$ are the real and imaginary parts of the 2D complex filter for orientation θ.

To reduce spectral leakage and preserve amplitude fidelity, each local phase signal is windowed using a Tukey window function with a cosine to constant ratio of 0.1. A fast Fourier transform is then applied to the phase signals over time for each pixel within a selected region of interest of the frame. To ensure reliability, only pixels with sufficient spatial gradients are used for analysis. The average frequency spectrum of the filtered pixels yields the dominant resonant frequencies of the bio-construct. These frequencies are used as indicators of the construct's dynamic behavior.

To further visualize the structural mode shapes, phase-based motion magnification is performed. Temporal bandpass filtering around the identified resonant frequencies is applied to the phase signal, and the resulting signals are amplified using a magnification factor. This amplified phase data is recombined with the original amplitude information to reconstruct a motion magnified video, revealing the operational deflection shapes (ODS) of the structure under vibration. These ODS visualizations provide valuable insights into the impact of embedded defects on the dynamic behavior of soft bioprinted constructs.

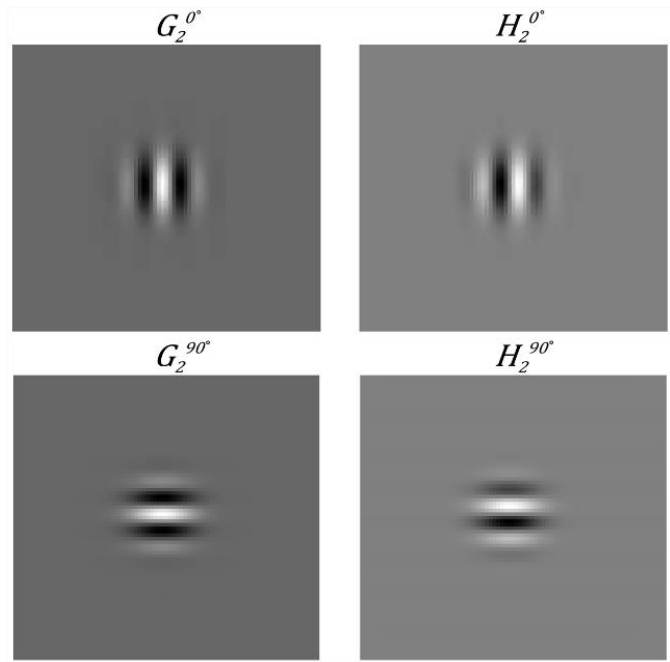

**Figure 3. 2D Gabor filter for two different orientations.**

**2.5 Finite Element Analysis**

To validate the experimental observations and investigate the modal characteristics of the bioprinted constructs under controlled conditions, finite element analysis was conducted using ANSYS Workbench 2024 R1 (Synopsys). The simulation model was developed to closely replicate the geometry, boundary conditions, and material behavior of the fabricated samples. The 3D models of both baseline and defective ear-shaped constructs were generated using SolidWorks (Dassault Systèmes) and subsequently imported into ANSYS for simulation. The printed geometries were derived directly from the G-Code slicing process to ensure dimensional consistency between simulated and physical constructs. Material properties for the hydrogel were defined based on the experimentally characterized formulation. The hydrogel was modeled as a linear elastic, isotropic material with Young's modulus of 9,500 Pa, density of 1,100 kg/m$^3$ and a Poisson's Ratio of 0.4, consistent with the behavior of the soft hydrogels used in the experimental analysis. For meshing, a quadratic element was chosen to improve solution accuracy, with an element size of 0.01 mm. A fixed boundary condition was applied at the base of the ear structure. The structure's natural frequencies and corresponding mode shapes were computed under free vibration conditions, simulating the response to an impulse-like excitation as observed from the experimental setup.

## 3. Results and Discussions

### 3.1 Reliability of the Measurements

To evaluate the reliability of the experimental setup and measurements, three defect-free samples were fabricated, and their frequency responses were obtained three times each using the high-speed camera and video-based vibration analysis. As shown in Figure 4, the fundamental resonant frequency corresponding to the flexural mode appeared at approximately 30 Hz, along with its harmonics, for all baseline samples. To extract the resonant frequencies, a Lorentzian curve was fitted to the frequency spectrum obtained from the measurements obtained in this study [35], [36], [37]. The resulting frequencies in Table 1, along with the small variability across the samples, demonstrate that the experimental setup yields reliable measurements and that the video-based motion estimation method effectively captures fine vibrational details in soft, translucent bioconstructs for non-invasive modal analysis.

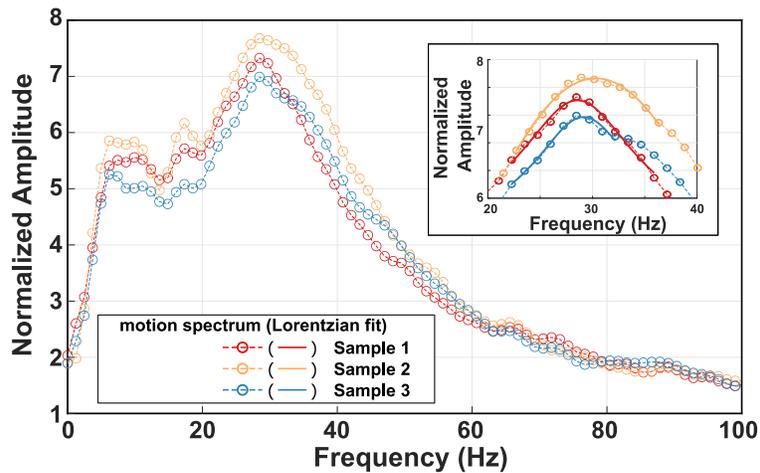

**Figure 4. Frequency response of baseline samples near the fundamental resonant frequency. The inset shows the Lorentzian fit (solid curves) to experimental data (circles) near the fundamental resonant frequency.**

**Table 1: Experimentally measured resonant frequencies of three baseline samples.**

| Fundamental Resonant Frequency, Hz | Sample 1 | Sample 2 | Sample 3 |
|---|---|---|---|
| **Mean** | 29.0 | 30.9 | 29.2 |
| **Standard Deviation** | 1.2 | 0.7 | 0.7 |

By applying temporal bandpass filtering to the local phase signals (from the video) within a frequency range that includes the fundamental resonant frequency and amplifying them, an enhanced visualization of the specific motion of interest, the operational deflection shape near the resonant frequency, can be obtained [34], [35]. This qualitatively represents the corresponding mode shape. Figure 5 represents a visualization of the mode shape of a baseline structure at the fundamental natural frequency obtained from finite element analysis. This mode shape was also obtained from the operational deflection shape of the baseline sample obtained through motion-magnified video (filtered at 29-31 Hz). The fundamental mode shape of the sample in this study reveals a flexural motion in which the helix part of the ear structure appears to sway back and forth. This sway-like motion indicates that the dynamic response is predominantly governed by variations in stiffness and mass distribution, similar to the swaying motion of a vertically standing beam structure.

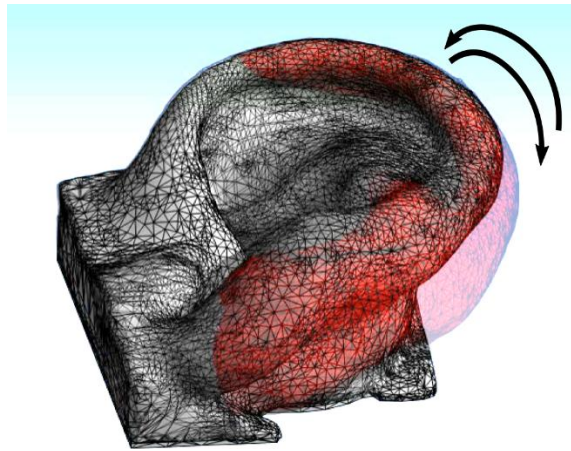

**Figure 5. Flexural motion in operational deflection shape is obtained from finite element analysis. This mode shape was also observed after magnifying the motions near the fundamental resonant frequency (29-31 Hz).**

## 3.2 Pressure-induced Defects

In extrusion-based bioprinting, maintaining consistent and uniform extrusion at the nozzle is essential for maintaining high print quality, structural integrity, and uniform layer deposition throughout the construct. However, defects such as filament thinning, discontinuities, or thickening can occur due to nozzle clogging, material viscosity variation, or irregularities in extrusion pressure, which can lead to pressure-induced defects [13], [14], [15]. This type of defect may result in irregular material deposition, weakened interlayer adhesion, or the formation of internal voids or inconsistent filament geometry in hydrogel-based bioprinted constructs [38], [39].

To investigate the effect of extrusion pressure variability on the dynamic response of bioprinted structures, we experimentally introduced pressure-induced defects by systematically varying the extrusion pressure of the bioprinter at different printing layers. Specifically, the pressure was altered at certain layers to induce volumetric inconsistencies in the printed constructs, allowing us to evaluate the effectiveness of the proposed defect detection method. For the defect-free (healthy) constructs, the extrusion pressure was set at 22 kPa throughout the print. In contrast, to generate defective samples, the extrusion pressure was increased from 22 kPa to 24 kPa and 26 kPa starting at layers corresponding to 25% and 50% of the total height, respectively, during the printing process, which resulted in four different types of defective samples (Figure 6). Three samples were printed for each defect condition, and each sample was measured three times. These changes are designed to emulate scenarios of sustained over-extrusion as a preliminary study, which can lead to material accumulation, poor interlayer fusion, and internal inconsistencies.

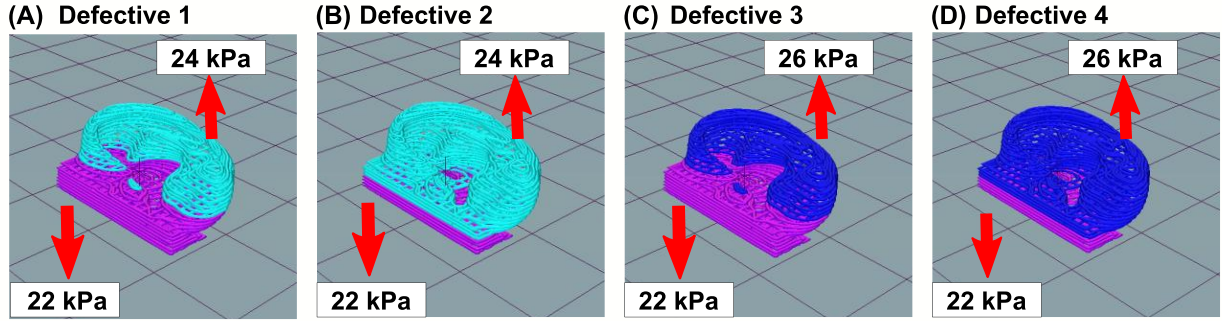

**Figure 6. Schematic representation of extrusion pressure variation used to induce pressure-related defects in bioprinted ear constructs: (A) Defective case 1 – the first 50% of the layers printed at 22 kPa followed by 24 kPa, (B) Defective case 2 – first 25% at 22 kPa and the rest at 24 kPa, (C) Defective case 3 – first 50% at 22 kPa and the rest at 26 kPa, and (D) Defective case 4 – first 25% at 22 kPa and the rest at 26 kPa.**

The frequency responses of the healthy and defective constructs, captured by the video-based vibrometry technique, are shown in Figure 7(A). Figure 7(B) compares the resonant frequencies across all groups, providing the mean and standard deviation with error bars. Here, it is clearly observed that the fundamental resonant frequency gradually shifts downward as the severity of the pressure defect increases. This result indicates that even minor deviations in extrusion parameters can measurably alter the dynamic response of bioprinted constructs, which can be effectively detected using the proposed vibration-based method.

In the defective cases, the base region was printed under the same extrusion pressure of 22 kPa as in the healthy conditions, so it may not show significant changes in stiffness. In contrast, the elevated extrusion pressure near the top portion may have increased the effective mass. Based on the well-established relation for resonant frequency $\omega_n = \sqrt{k_e/m_e}$ (where $k_e$ and $m_e$ represent the effective stiffness and mass of the structure, respectively), this observation implies that the increase in effective mass contributed to a reduction in resonant frequency, consistent with theoretical expectations.

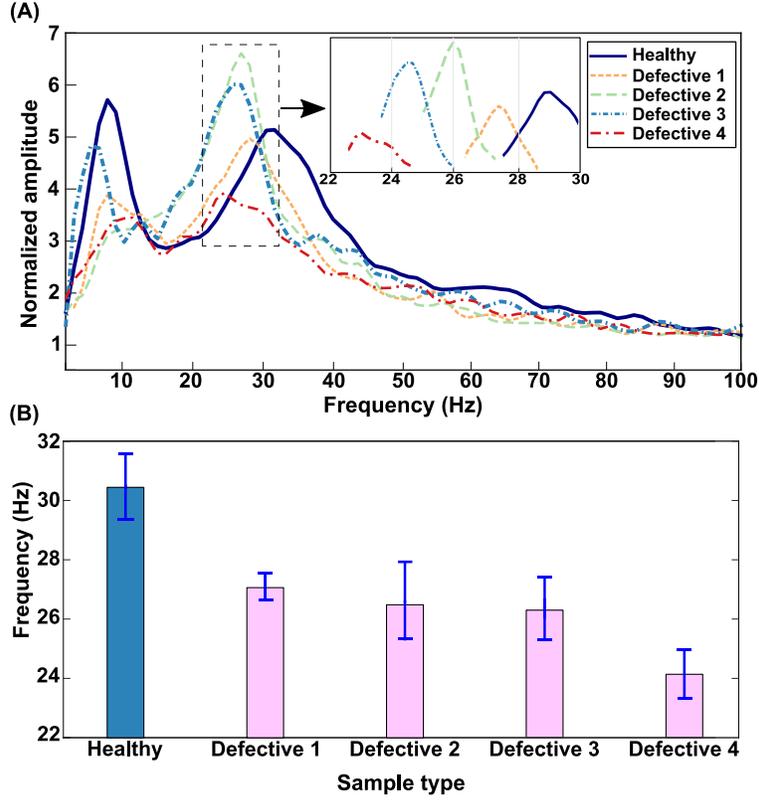

**Figure 7. Comparison of resonant frequencies between baseline and pressure-induced defective samples. (A) Frequency response and (B) Fundamental resonant frequencies for each case. Error bars indicate one standard deviation.**

To further investigate the impact of extrusion pressure variation on the dynamic behavior of bioprinted constructs, the root mean square deviation (RMSD) and mean amplitude percentage deviation (MAPD) were employed as damage indices to quantitatively compare the frequency responses of healthy and defective samples. The RMSD and MAPD are two data-driven metrics that have been widely adopted in vibration-based structural health monitoring [22], [25], [40], [41], valued for its capacity to quantify subtle deviations between baseline and damaged conditions. The RMSD and MAPD are obtained as follows:

$$RMSD(\%) = \sqrt{\frac{\sum_{i=1}^{i=N}(Y_i-X_i)^2}{\sum_{i=1}^{i=N} X_i^2}} \times 100 \qquad (2)$$

$$MAPD(\%) = \frac{1}{N}\sum_{i=1}^{N}\left|\frac{Y_i-X_i}{X_i}\right| \times 100 \qquad (3)$$

where $X_i$ denotes the mean amplitude of the baseline samples and $Y_i$ denote the amplitude of the target samples, within the specified frequency range that contains $N$ data points.

Figure 8 respectively illustrate the RMSD and MAPD values calculated over a frequency range (11–50 Hz) encompassing the fundamental resonant frequencies for the healthy and defective samples across four defect cases. The data-driven metrics (RMSD and MAPD) show significantly higher value for all defective cases compared to the corresponding healthy baseline, indicating large deviations in the spectral amplitude caused by extrusion pressure variations.

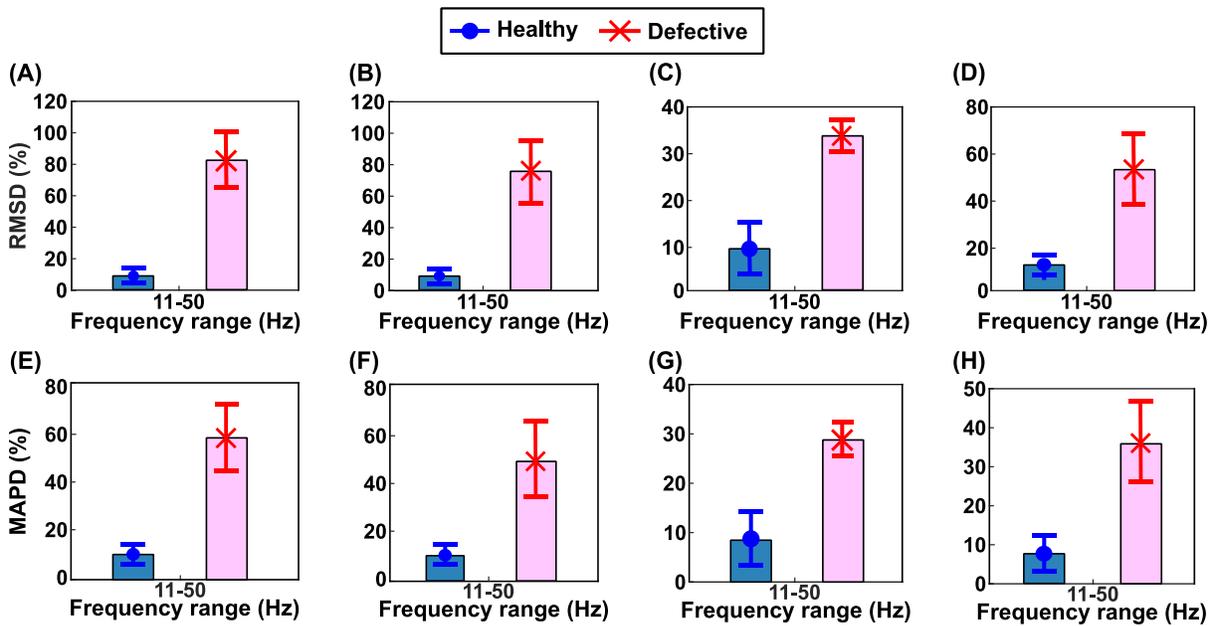

**Figure 8. Comparison of RMSD and MAPD of amplitudes between healthy and pressure-induced defective samples over the frequency range (11–50 Hz). (A–D) present the RMSD and (E-H) represent MAPD comparison respectively for four different pressure-induced defective printing conditions: (A, E) healthy vs defect case 1 – 22 kPa (50%) and 24 kPa (50%), (B, F) healthy vs defect case 2 – 22 kPa (25%) and 24 kPa (75%), (C, G) healthy vs defect case 3 – 22 kPa (50%) and 26 kPa (50%), (D, H) healthy vs defect case 4 – 22 kPa (25%) and 26 kPa (75%).**

To localize and further quantify the vibrational characteristics introduced by extrusion pressure variations, the above frequency spectrum was segmented into four distinct bands, and data-driven metric values were computed within each band for all sample groups. The RMSD- and MAPD-based comparisons between healthy and pressure-defective constructs across these frequency ranges for the four different defect scenarios are presented in Figure 9. Across all defect types, a consistent trend can be observed. Both data-driven metric values peak most prominently within the 21–30 Hz range, which closely aligns with the fundamental resonant frequencies of the bioconstructs. The increased metric in this range indicates that pressure-induced structural defects have the most significant effect on the vibrational response near the primary resonant frequency [22], [25]. Overall, the experimental investigation results confirm the effectiveness of monitoring resonant frequency shifts and data-driven metrics in detecting pressure-induced defects in bioprinted constructs through the proposed video-based vibration analysis.

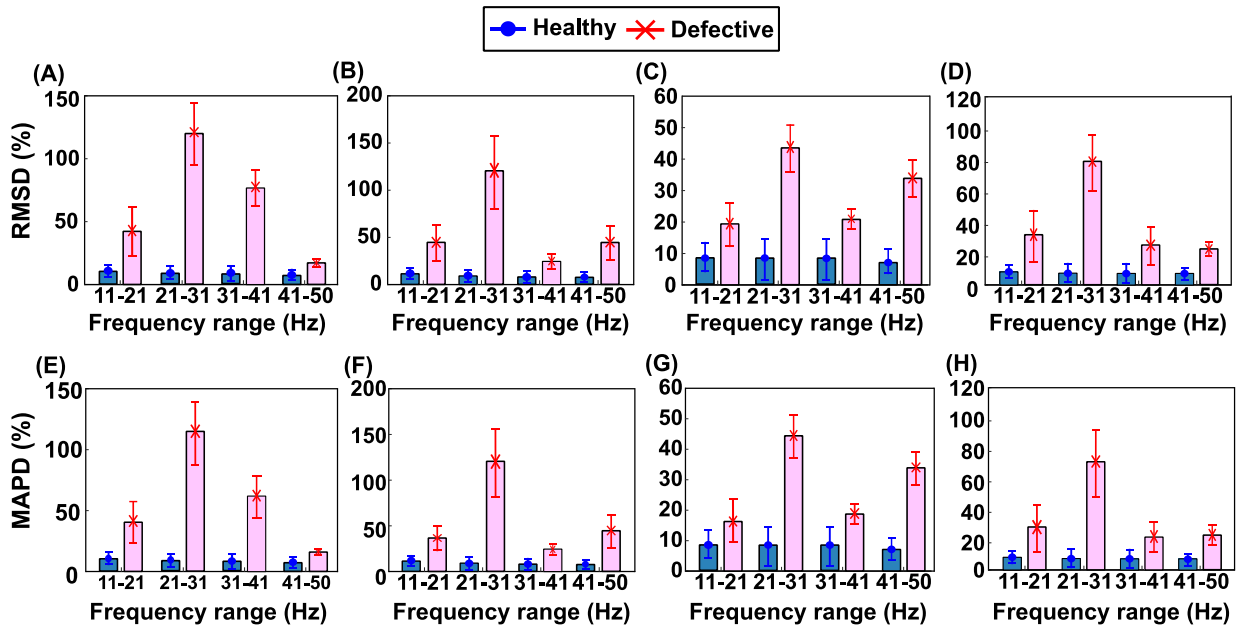

Figure 9. Comparison of RMSD and MAPD of amplitudes between healthy and pressure-induced defective samples across four frequency bands from 11–50 Hz. (A–D) present the RMSD and (E-H) represent MAPD comparison respectively for four different pressure-induced defective printing conditions: (A, E) healthy vs defect case 1 – 22 kPa (50%) and 24 kPa (50%), (B, F) healthy vs defect case 2 – 22 kPa (25%) and 24 kPa (75%), (C, G) healthy vs defect case 3 – 22 kPa (50%) and 26 kPa (50%), (D, H) healthy vs defect case 4 – 22 kPa (25%) and 26 kPa (75%).

## 3.3 Geometric Defects

Geometric defects are among the most frequently encountered issues in extrusion-based bioprinting and can significantly compromise the fidelity and mechanical functionality of bioconstructs. These defects often arise from transient nozzle obstructions, inconsistent extrusion, or air bubble entrapment, which results in incomplete material deposition or void formation within the printed structure. Those defects can alter the mechanical response of the construct and undermine its performance in tissue engineering applications.

In this study, the geometric defect was introduced into the ear-shaped sample as a controlled hole (Figure 10). Geometric defects are introduced in the CAD design by incorporating circular voids of three different diameters (2 mm, 3 mm, and 4 mm) near the ear lobule region. Among the three defect sizes, the 2 mm void was completely concealed beneath successive layers due to bioink spreading and fusion between layers, while the 3 mm defect presented a partially visible irregularity. These observations emphasize the challenge of visually identifying small voids in bioconstructs, which are often made of translucent hydrogels, and highlight the necessity of structural health monitoring techniques for detecting subsurface anomalies.

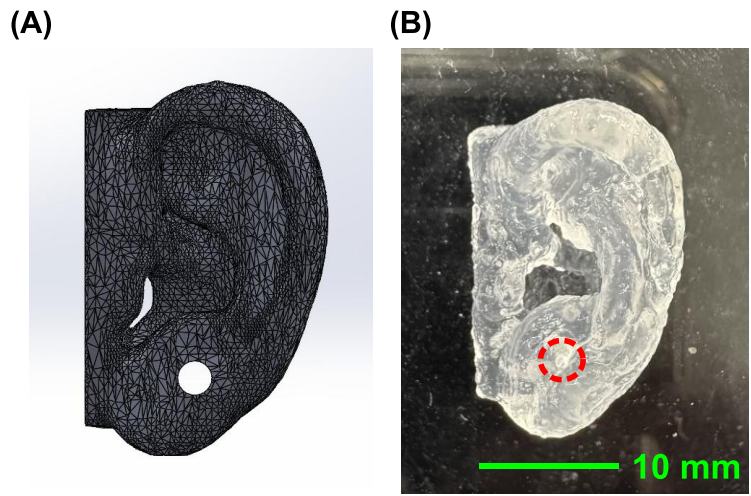

**Figure 10. Geometrically defective bioprinted ear samples: (A) CAD model, and (B) corresponding 3D bioprinted model. The geometric defect (3 mm diameter hole) is highlighted with a red dashed circle in (B).**

Figure 11(A) presents the frequency response of the healthy baseline and geometrically defective samples obtained by using the video-based vibration analysis method. The fundamental resonant frequencies for healthy and all defective cases are compared in Figure 11(B) with error bars indicating one standard deviation. The baseline samples exhibit the lowest resonant frequency, while the group with the largest defect shows the highest. This trend can be attributed to the reduction in the effective mass caused by the void, which effectively alters the stiffness-to-mass ratio and thereby increases the resonant frequency of the structure. To validate the experimental findings, modal analysis was performed using finite element simulations, replicating the geometric structures and material properties of the bioprinted samples. The numerical simulation results (dashed curve in Figure 11(B)) show a good qualitative and quantitative agreement with the experimental findings, demonstrating that the proposed video-based vibration analysis can reliably detect geometric voids in soft bioprinted constructs.

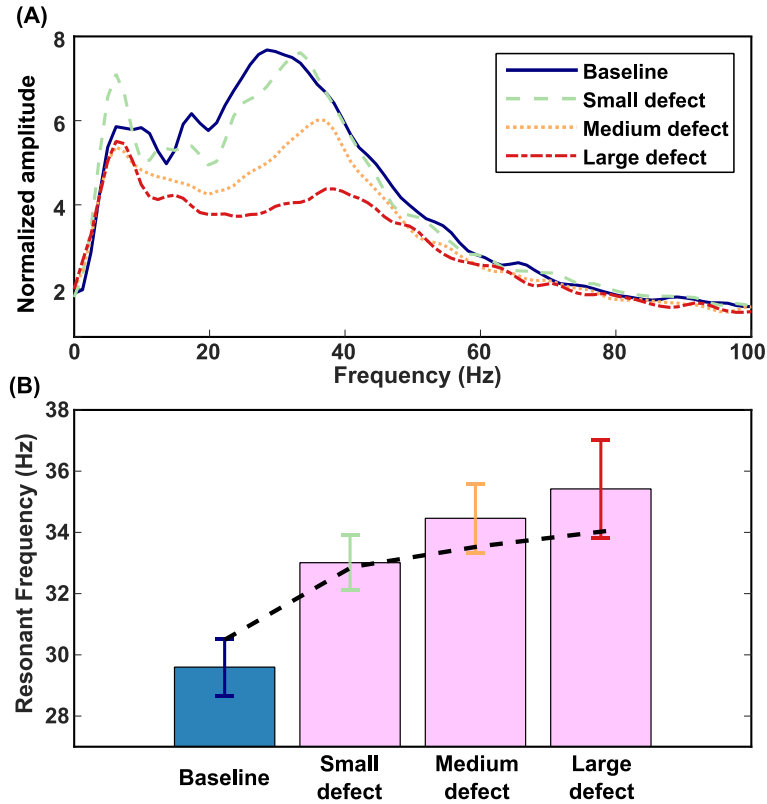

**Figure 11. (A) Comparison of experimentally obtained frequency responses between baseline and geometric defective samples. (B) Experimentally obtained fundamental resonant frequencies of baseline and geometric defective samples. Error bars denote one standard deviation. The dashed curve indicates numerically obtained fundamental resonant frequencies via finite element analysis.**

To investigate how geometric voids influence the vibrational behavior of bioprinted constructs, the data-driven metrics, RMSD and MAPD, were analyzed over a broad frequency range (11–50 Hz) for healthy samples and samples with three levels of intentionally induced geometric defects (Figure 12). As the severity of the geometric defect increases, both RMSD and MAPD values gradually increase, quantitatively reflecting the extent to which the frequency responses of defective samples deviate from those of the healthy ones.

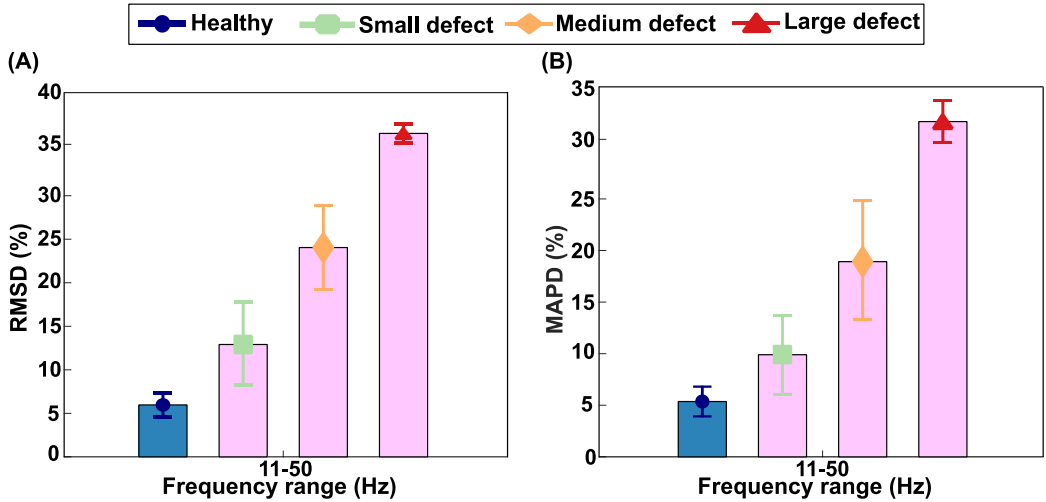

**Figure 12. Comparison of amplitude deviation between healthy and geometrically defective samples over the frequency range (11–50 Hz). (A) RMSD % (B) MAPD % are presented for healthy samples and three levels of geometric defects (small, medium, and large).**

To further analyze the localized vibrational effect of internal geometric voids, Figure 13 presents the multi-zone RMSD and MAPD values for healthy and defective samples across four frequency bands ranging from 11 Hz to 50 Hz. As the defect size increases, both data-driven metrics from healthy baselines become more prominent across all zones. Even at smaller defect sizes, an increase in RMSD and MAPD values within the 21–41 Hz range indicates notable deviations in the frequency response, yet the magnitude of this deviation remains relatively small. As the defect size increases, however, the data-driven metric values within the 21–31 Hz range become remarkably larger, clearly distinguishing more severe damage. This highlights the resonant frequency range as the most sensitive to geometric defects. Overall, the multi-range analysis demonstrates that evaluating distinct frequency bands can more effectively isolate the dynamic characteristics associated with specific damage levels.

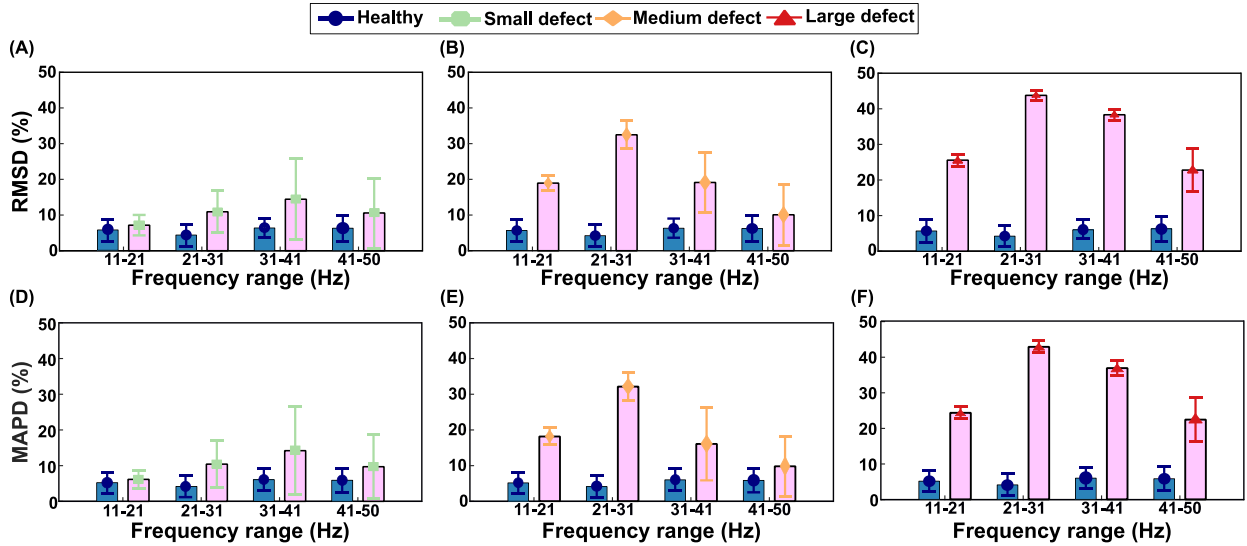

**Figure 13. Comparison of RMSD and MAPD of amplitudes between healthy and geometrically defective samples across four frequency bands from 11–50 Hz. (A–C) present the RMSD and (D-F) represent MAPD comparison respectively for three different geometrically defective samples.**

## 3.4 Interlayer Defects

Interlayer defects are structural anomalies that arise from incomplete bonding or material discontinuity between successive layers during the bioprinting process. These defects are particularly detrimental in bioconstructs, where weak adhesion between layers can significantly impair mechanical integrity [42], [43]. In extrusion-based bioprinting, interlayer defects may result from irregular material flow, inconsistent pressure, or transient skipping in the deposition path from G-Code. As shown in Figure 14(A), such defects often occur in regions with overhangs [44], [45]. In this study, to systematically introduce interlayer defects in experiments, which often occur unpredictably and at random, the print path in the G-Code was modified so that, at specific layers, a single path near the layer's edge was skipped. Figure 14(B, C) illustrates the CAD models of such cases for the ear-shaped sample, where the print path along the front and rear edges of the ear was intentionally skipped at one of the layers that correspond to the helix part of an ear with an overhang-like geometry. After fabrication, neither of the defects was externally visible because of layer fusion, making these practical cases for subsurface detection methods.

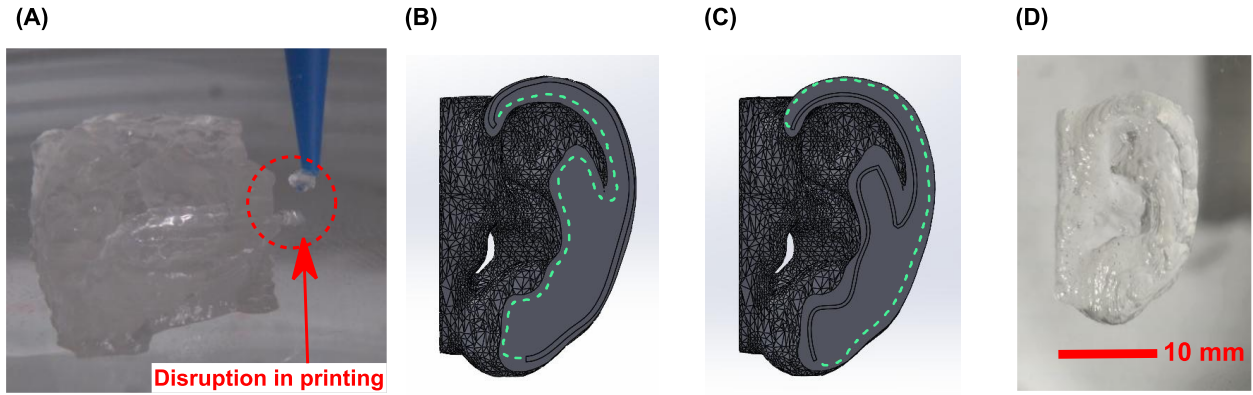

**Figure 14.** Illustration of interlayer defect formation: **(A)** An image captured during printing, highlighting interlayer defects with disruption in printing (red circle), **(B, C)** CAD model representing the front and rear type of interlayer defects (missing path highlighted with dashed line), and **(D)** Final printed ear sample with a rear-edge interlayer defect.

Figure 15(A) compares the frequency responses of two defective cases with that of the baseline. As shown in Figure 15(B), the interlayer defect along the rear edge increases the resonant frequency, whereas the defect along the front edge lowers it. As discussed in Section 3.2, when the sway-like mode shape corresponding to the fundamental resonant frequency is considered analogous to a vertically standing beam structure, the rear-edge interlayer defect appears to reduce the effective mass, thereby increasing the resonant frequency (Figure 15(B)). In contrast, the front-edge defect may reduce the effective stiffness near the base, resulting in a decrease in the resonant frequency (Figure 15(C)). This trend is further supported by the finite element–based numerical analysis, which provided quantitatively and qualitatively similar results (dashed curves in Figure 15(B, C)).

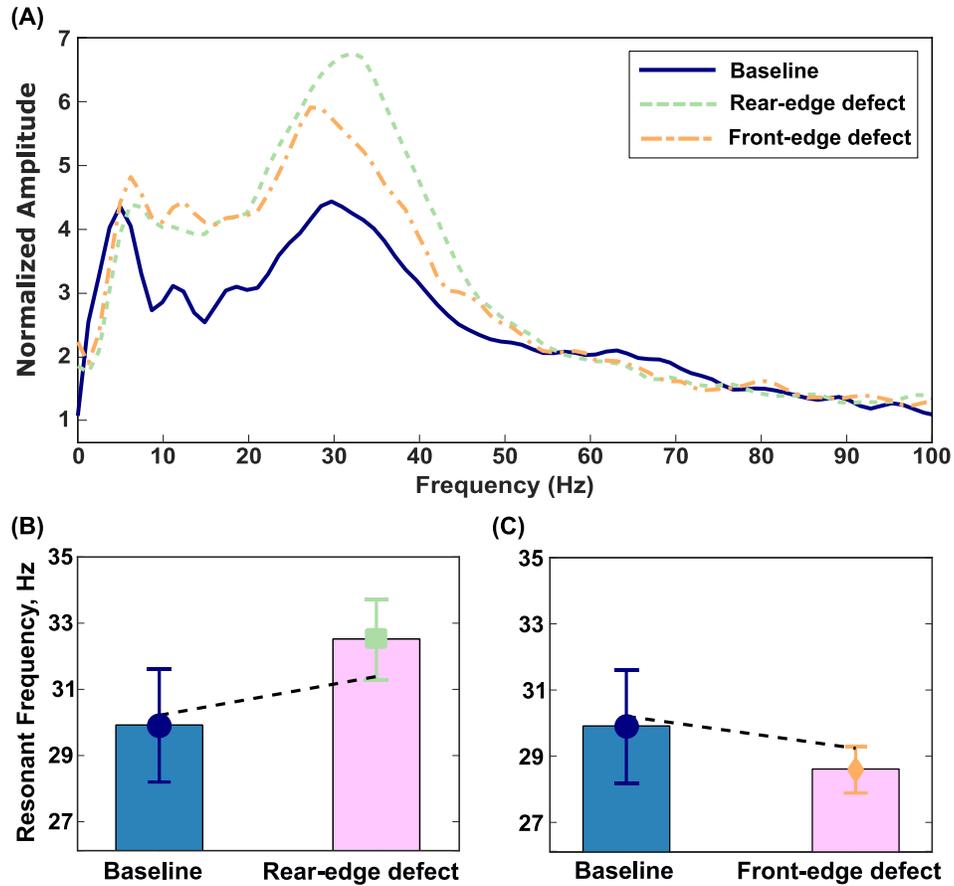

Figure 15. Comparison of resonant frequencies between baseline and Interlayer defective samples. (A) Frequency response and (B) Fundamental resonant frequencies for each case. Error bars indicate one standard deviation. The curve indicates numerically obtained fundamental resonant frequencies via finite element analysis.

Figure 16 presents the data-driven metrics (RMSD and MAPD) computed over the full frequency range of 11–50 Hz to compare the healthy and interlayer-defective bioprinted samples. Compared to the relatively small resonant frequency shifts (Figure 15(A)), the data-driven metrics reveal these changes much more distinctly (Figure 16). Compared to the healthy sample, both the RMSD and MAPD indices show a substantial increase in the rear-edge defect, while front-edge defect also exhibits higher values than the healthy sample but lower than the rear-edge case. This trend is consistent with the degree of frequency shift observed in the frequency response (Figure 15(B, C)). This progressive increase in deviation from the healthy to the interlayer-defective cases highlights the sensitivity of both RMSD and MAPD to structural anomalies arising from layer-wise printing inconsistencies.

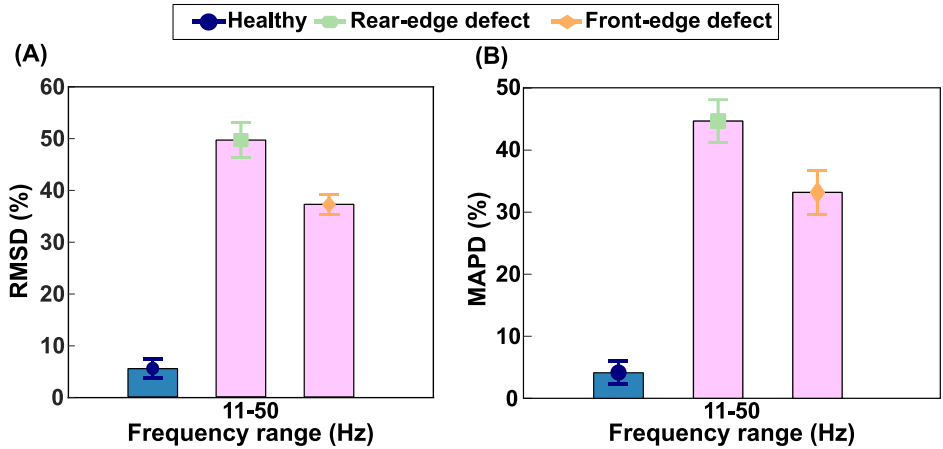

**Figure 16.** Comparison of (A) RMSD and (B) MAPD values between healthy samples and two types of interlayer defective samples (missing the front and rear part of the 14th layer, respectively) across full frequency bands (11-50 Hz).

Similar to the previous section, when the data-driven metrics were computed over each sub frequency range, an overall increase in these metrics was observed across the entire frequency domain due to the damages (Fig. 17). The most pronounced increase occurred in the range of 21–41 Hz, which includes the fundamental resonant frequency. This trend, which is consistent with results obtained for other defect types, further confirms that vibration-based analysis exhibits high sensitivity to variations induced by embedded defects, particularly in the vicinity of resonance. Overall, these results demonstrate that subtle interlayer inconsistencies within the structure, which is challenging to detect through visual inspection, can be identified by monitoring the overall dynamic behavior using video-based vibrometry.

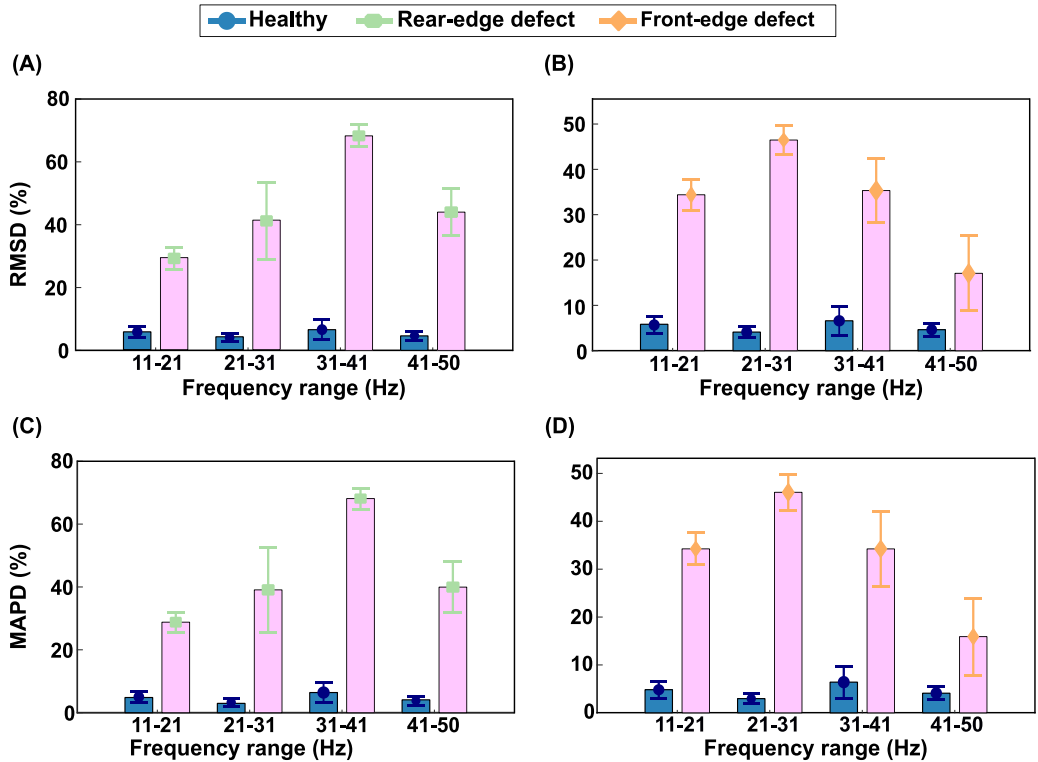

**Figure 17. Comparison of RMSD and MAPD values between healthy and two types of interlayer defects. (A-B) represent comparison of RMSD between healthy sample and the rear-edge defect, and between the healthy and the front-edge defect, respectively. (C-D) represent comparison of MAPD between the healthy sample and the rear-edge defect, and between the healthy sample and the front-edge defect, respectively.**

## 4. Conclusion

In summary, this research presents a robust and non-contact approach for non-destructively evaluating structural integrity in soft bioprinted constructs by leveraging video-based vibration analysis. Using phase-based motion estimation techniques applied to high-speed video data, the study successfully captured the dynamic response of 3D printed ear-shaped scaffolds fabricated via extrusion-based bioprinting. The approach enables the identification of surface and embedded defects by analyzing changes in the frequency spectra through monitoring resonant frequency shifts and data-driven metrics. Three common categories of printing defects, such as extrusion pressure irregularities, geometric voids, and interlayer discontinuities were systematically introduced and analyzed. The numerical and experimental investigation results demonstrate the effectiveness of the proposed video-based vibration analysis in detecting various types of structural defects in bioprinted constructs. Overall, the proposed method demonstrates strong potential as a non-destructive, non-invasive approach for structural health monitoring in bioprinting. By enabling early detections of various defect types without contact or embedded sensors, this method contributes to advancing print quality and reliability in bio-additive manufacturing.

## 5. Acknowledgements

This research is supported by the National Science Foundation under Award No. 2301948 and by the faculty research seed grant from the College of Engineering and Computing at Georgia Southern University.